%Paper: hep-th/9402145
%From: Benjamin Enriquez <enriquez@orphee.polytechnique.fr>
%Date: Fri, 25 Feb 1994 18:07:05 +0100 (MET)

\magnification\magstep1
\hsize 15.5truecm
% TO SAVE AS texnot.tex
 at 17.28truept
 at 14.4truept
 at 12truept
 at 10.95truept
 at 17.28truept
%\font\smcap=cmcsc10 at 10truept
 at 10truept
% Un "format" vraiment minimal pour "chemin" tape
% par Marie-Jo
%\input amstex
%\documentstyle{amsppt}

\def\title#1{%
\vskip0pt plus.3\vsize\penalty-100%
\vskip0pt plus-.3\vsize\bigskip\vskip\parskip%
\bigbreak\bigbreak\centerline{\bf #1}\bigskip%
}

\def\chapter#1#2{\vfill\eject%		chapitre avec 1 ligne
\centerline{\bf Chapter #1}%		de titre
\vskip 6truept%
\centerline{\bf #2}%
\vskip 2 true cm}

\def\section#1#2{%
\def\\{#2}%
\vskip0pt plus.3\vsize\penalty-100%
\vskip0pt plus-.3\vsize\bigskip\vskip\parskip%
\par\noindent{\bf #1\hskip 6truept%
\ifx\empty\\{\relax}\else{\bf #2\smallskip}\fi}}

\def\subsection#1#2{%
\def\\{#2}%
\vskip0pt plus.3\vsize\penalty-20%
\vskip0pt plus-.3\vsize\medskip\vskip\parskip%
\def\TEST{#1}%
\noindent{\ifx\TEST\empty\relax\else\bf #1\hskip 6truept\fi%
\ifx\empty\\{\relax}\else{#2\smallskip}\fi}}

\def\proclaim#1{\medbreak\begingroup\noindent{\bf #1.---}\enspace\sl}

\def\endproclaim{\endgroup\par\medbreak}

\def\qqbox#1{\quad\hbox{#1}\quad}

%pour laisser de la place pour 1 figure:
%\midinsert\vskip 8 true cm
%\centerline{\sl Fig. 1}
%\endinsert

%Fig. avec legende au-dessus du numero.
\def\comfig#1#2\par{
\medskip
\centerline{\hbox{\hsize=10cm\eightpoint\baselineskip=10pt
\vbox{\noindent #1}}}\par\centerline{ Figure #2}}

%Fig. avec legende au-dessous du numero.
\def\figcom#1#2\par{
\medskip
\centerline
{Figure #1}
\par\centerline{\hbox{\hsize=10cm\eightpoint\baselineskip=10pt
\vbox{\noindent #2}}}}
% Pour les fins demo, un taureau carre
\def\bull{~\vrule height .9ex width .8ex depth -.1ex}

\def\adresse#1{%
\bigskip\hfill\hbox{\vbox{%
\hbox{#1}\hbox{Centre de Math\'ematiques}\hbox{Ecole Polytechnique}%
\hbox{F-91128 Palaiseau Cedex (France)}\hbox{\strut}%
\hbox{``U.A. au C.N.R.S. n$^{\circ}$169''}}}}

%\adresse{Tartanpium}

%Fig. avec legende au-dessus du numero.
\def\comfig#1#2\par{
\medskip
\centerline{\hbox{\hsize=10cm\eightpoint\baselineskip=10pt
\vbox{\noindent{\sl  #1}}}}\par\centerline{{\bf Figure #2}}}

%Fig. avec legende au-dessous du numero.
\def\figcom#1#2\par{
\medskip
\centerline
{{\bf Figure #1}}
\par\centerline{\hbox{\hsize=10cm\eightpoint\baselineskip=10pt
\vbox{\noindent{\sl  #2}}}}}

\def\em{\sl}

\def\\{\hfill\break}

\def\bibitem{\item}

\long\def\adresse#1{%
\leftskip=0truecm%
\vskip 3truecm%
\hbox{\hskip 10.5truecm{\hsize=7.5truecm\vbox{%
\def\cr{\par\noindent}\noindent#1}}}}

%Le carre de fin de demonstration idem que CQFD ou QED
\def\bull{~\vrule height .9ex width .8ex depth -.1ex}

% TO SAVE AS Macros.tex
\def\a{\alpha}\def\b{\beta}\def\d{\delta}
\def\e{\varepsilon}\def\t{\theta}
\def\la{\lambda}
\def\Si{\Sigma}
\def\f{\varphi}

\def\CC{{\bf C}}

\def\ZZ{{\bf Z}}

\def\tr{\mathop{\rm tr}\limits}

%pour pouvoir ecrire en-dessous
%pour pouvoir ecrire en-dessous
%pour pouvoir ecrire en-dessous
%pour pouvoir ecrire en-dessous
%pour pouvoir ecrire en-dessous

\def\la{\lambda}

\def\sqr#1#2{{\vcenter{\hrule height.#2pt%
\hbox{\vrule width.#2pt height#1pt\kern#1pt%
\vrule width.#2pt}%
\hrule height.#2pt}}}

%Premiere section

\def\nd{\noindent}

\def\wh{\widehat}
\def\difpol{differential polynomials}
\def\kdv{KdV}
\def\np{\widehat n_{+}}
\def\wsl{\widehat{s\ell}}
\def\pol{polynomials}

\overfullrule=0pt

\centerline{\bf
Quantum principal
 commutative subalgebra in the
}
\centerline{\bf nilpotent part of $U_q\widehat{s\ell}_2$ and lattice KdV
variables }

\medskip
\centerline{\rm B. Enriquez}

\medskip

\noindent
{\bf Abstract.} {\em
We propose a quantum lattice version of Feigin and E. Frenkel's constructions,
identifying the KdV differential polynomials with functions on a homogeneous
space under the nilpotent part of $\widehat{s\ell}_2$. We construct an action
of the nilpotent part $U_q\widehat n_+$ of $U_q\widehat{s\ell}_2$ on their
lattice counterparts, and embed the lattice variables in a
$U_q\widehat n_+$-module, coinduced from a quantum version of the principal
commutative subalgebra, which is defined using the identification of
$U_q\widehat n_+$ with its coordinate algebra.  }

\medskip
\noindent

\section{Introduction.}{}

In [FF1], [FF2], Feigin and E. Frenkel propose a new approach to the
generalized KdV hierarchies.
They construct an action of the nilpotent part $\wh n_{+}$ of the affine
 algebra $\wh g$ on \difpol \ in the Miura fields, connected to the action of
screening operators. This enables them to consider these \difpol \  as
functions on a homogeneous space of $\wh n_{+}$, and to interpret in this way
the KdV flows. They also suggest that analogous constructions should hold for
the quantum \kdv \ equations.

In this work we propose a quantum lattice version of part of these
constructions. Following ideas of lattice $W$-algebras, we replace the \difpol
\ by an algebra of $q$-commuting variables, set on a half-infinite line. The
analogue of the action of [FF1] is then an action of the nilpotent part
$U_{q}\wh n_{+}$ of the quantum affine algebra $U_{q}\wh{s\ell}_{2}$. Recall
that the
homogeneous space occuring in [FF1] is $\wh N_{+}/A$, where $\wh N_{+}$ and
$A$ are the groups corresponding to $\np$ and its principal commutative
subalgebra
$a$. A natural question is then what the analogue of $a$ is in the quantum
situation.

We construct a quantum analogue of $a$ in the following way~: we use an
isomorphism of $U_{q}\wh b_{+}$ with the coordinate ring $\CC[\wh B_{+}]_{q}$
([Dr], [LSS]) and transport in the first algebra a twisted version of the
well-known commutative family res $d\la\la^{k} \tr\, T(\la)$. We prove that
this subalgebra of $U_{q}\wh b_{+}$ gives $Ua$ for $q=1$. This proof uses
characterizations of these algebras as centralizers of one element.

Using a realization of the coordinate ring $\CC [\wh B_{+}]_{q}$ in
$q$-commuting variables, due to Volkov, we find explicit expressions for the
representation of $U_{q}a$ in operators on the half line. A symmetry argument
then shows the analogue of the result of Feigin and Frenkel: injection of the
lattice variables in a module coinduced from $U_{q}a$ to $U_{q}\widehat b_{+}$.

\section{1.}{The approach of Feigin and E. Frenkel.}
Let us recall briefly the part of [FF1] we will be concerned with (in the
$\wh{s\ell}_{2}$ case). Let $\phi$ be the free field on $S^{1}$
$\{\phi(x),\phi(y)\}=\d'(x-y)$, and $\f'=\phi$. There is an action of the upper
 nilpotent part of $\wh{s\ell}_{2}$ on the algebra
$\CC[\phi(x),\phi'(x),\dots]$ of polynomials in $\phi(x),\phi(x),\dots,$ given
by
$Q_{+}P(\phi(x),\phi'(x),\dots)=e^{-\f(x)}\{\int_{S^{1}}e^{\f},P(\phi(x),
\phi'(x),\dots)\}$
 and
$Q_{-}P(\phi(x),\phi'(x),\dots)=e^{\f(x)}\{\int_{S^{1}}e^{-\f},P(\phi(x),
\phi'(x),\dots)\}$~;
 $Q_{+}$ and $Q_{-}$ are the usual generators of $\wh n_{+}\subset
\wh{s\ell}_{2}$, satisfying the analogues of Serre relations.

There is a duality between $U\wh n_{+}$ and $\CC[\phi(x),\phi'(x),\dots]$,
given by
$$
\eqalign{
U\wh n_{+} &\times \CC[\phi(x),\phi'(x),\cdots ] \rightarrow \CC\cr
T &\times P\longmapsto \e(TP)=(TP)(\phi(x)=0,\phi'(x)=0,\dots).\cr}
$$

\nd
Here $\e$ is the operation of suppression of all non constant terms in a given
differential polynomial.

\nd
 Let $a\subset \wh n_{+}$ be the principal commutative subalgebra, spanned by
$Q_{+}+Q_{-}\, ,\, [Q_{+}-Q_{-},h(i)]$, $i\geq 1$, where $h(i)$ are inductively
defined  by $h(1)=[Q_{+},Q_{-}]$, $h(i+1)=[Q_{+},[Q_{-},h(i)]]$. Then
$\e(xP)=0$, if $x\in a$. The pairing thus factors through a pairing $(\CC
\mathop{\otimes}_{Ua}U\wh n_{+}
)\times\CC[\phi(x),\phi'(x),\dots]\to\CC$~; it
enables to identify $\CC[\phi(x),\phi'(x),\dots]$ with $\CC[\wh N_{+}/A]$ as
$\wh n_{+}$-module ($\wh N_{+}$ and $A$ being the groups corresponding to $\wh
n_{+}$ and $a$).

\section{2.}{The lattice setting.}

Let us consider variables $x_{i}$, $i\leq 0$, satisfying the relations
$x_{i}x_{j}=q x_{j}x_{i}$ if $i<j$~; they are thought of as analogues of
variables $e^{\f(-i)}$ and polynomials $\mathop{\prod}_{i\leq 0}x_{i}^{\a_{i}}$
with $\mathop{\sum}_{i\leq 0}\a_{i}=0$ as analogues of the differential
polynomials in
$\phi(x),\phi'(x),\dots,$ on the half infinite lattice $i\leq 0$, $i$ integer
(the point 0 of this lattice corresponds to $x$ in the continuous approach.)

\nd
On the space $\CC[x_{i}x_{0}^{-1}]$ of degree zero \pol , we define the
operators $Q_{+},Q_{-}$ and $K$ by
$$
Q_{+}P=[\sum_{i<0}x_{i},P]x_{0}^{-1}\, ,\,
Q_{-}P=[\sum_{i<0}x_{i},P]x_{0}\, ,\, KP=x_{0}P x_{0}^{-1}\, .
$$
\proclaim{Lemma 1} The operators $Q_{+}$, $Q_{-}$ satisfy the
$q$-Serre-Chevalley relations
$$
Q^{3}_{\pm}Q_{\mp}-(q^{2}+1+q^{-2})Q^{2}_{\pm}Q_{\mp}Q_{\pm}+(q^{2}+1+q^{-2})
Q_{\pm}Q_{\mp}Q^{2}_{\pm}-Q_{\mp}Q^{3}_{\pm}=0,
$$
and the relations $K Q_{\pm}=q^{\mp 1}Q_{\pm}K$. So they define an action of
$U_{q}\wh b_{+}\subset U_{q}\wh{s\ell_{2}}$ on $\CC[x_{i}x_{0}^{-1}]$ (the
level of $U_{q}\widehat{s\ell}_{2}$ is taken to be zero).
\endproclaim

\nd
{\it Proof.} We have
$$
Q_{+}\left(\prod_{i\leq
0}x_{i}^{\a_{i}}\right)=\sum_{j<0}\left(q^{-\sum_{s<j}\a_{s}}-q^{\sum_{s\leq
j}\a_{s}}\right)\prod_{i\leq 0}x_{i}^{\a_{i}+\d_{ij}}x^{-1}_{0}
$$
and
$$
Q_{-}\left(\prod_{i\leq
0}x_{i}^{\a_{i}}\right)=\sum_{j<0}\left(q^{\sum_{s<j}\a_{s}}-q^{-\sum_{s\leq
j}\a_{s}}\right)\prod_{i\leq 0}x_{i}^{\a_{i}-\d_{ij}}x_{0},
$$
if $\mathop{\sum}\limits_{i\leq 0}\a_{i}=0$

\nd
(the products are written with lower indices at the left, e.g.
$\mathop{\prod}_{i\leq 0}x_{i}^{\a_{i}}=\dots x_{n}^{\a_{n}}\dots
x_{0}^{\a_{0}}$). Let us associate to $\mathop{\prod}_{i\leq 0}x_{i}^{\a_{i}}$
the element $e^{\mathop{\sum}_{i< 0}\a_{i}\xi_{i}}$ in the (commutative)
algebra $\CC[e^{\pm \xi_{i}},\, i<0]$. In this representation, $Q_{\pm}$ can be
written
$$
Q_{\pm}=\sum_{j<0}\,
e^{\pm\xi_{j}}\left(q^{\mp\mathop{\sum}\limits_{s<j}{\partial\over\partial
\xi_{s}}}-q^{\pm\mathop{\sum}\limits_{s\leq j} {\partial\over\partial
\xi_{s}}}\right)\,
{}.
$$
Pose
$$
\t_{j}=e^{\xi_{j}}q^{\mathop{\sum}\limits_{s\leq j}{\partial\over\partial
\xi_{s}}}\, , \, \,\t^{-}_{j}=e^{-\xi_{j}}q^{-\sum_{s\leq
j}{\partial\over\partial\xi_{s}}}\, , \, \,
\t'_{j}=e^{\xi_{j}}q^{-\mathop{\sum}\limits_{s<j}{\partial\over\partial
\xi_{s}}}\, ,\, \, \t'^{-}_{j}=\t'^{-1}_{j}\, .
$$
Then
$$
Q_{+}=-\sum_{j<0} \t_{j}+\sum_{j<0} \t'_{j}\, \, ,\, \, Q_{-}=
-\sum_{j<0} \t_{j}^{-} +\sum_{j<0} \t'^{-1}_{j}\, .
$$
Remark that if $j>k$, $\t_{j}\t_{k}=q\t_{k}\t_{j}$,
$\t_{k}\t'_{k'}=q\t'_{k'}\t_{k}$, for all $k$ and $k'$, and $\t'_{k'}\t'_{j'}=q
\t'_{j'}\t'_{k'}$ if $k'<j'$. The two first relations can then be deduced from
the following result ([F], [KP])~:
\proclaim{Lemma 2 {\rm (B. Feigin)}} If $s^{\pm}_{i}$, $i\in\ZZ$ are variables
such that for $i<j$, $s^{\e}_{i}s^{\e'}_{j}=q^{\e \e'}s^{\e'}_{j}s^{\e}_{i}$,
$\e,\e'=\pm 1$, then $s^{\pm}= \mathop{\sum}\limits_{i\in\ZZ}s^{\pm}_{i}$
satisfy the $q$-Serre relations of $U_{q}\wsl_{2}$.
\endproclaim

\nd
{\it Proof.} (Note that we may have only a finite number of non vanishing
$s^{\pm}_{i}$.) Iterated application of the coproduct of $U_{q}\np$ gives an
algebra morphism $U_{q}\np\to (U_{q}\np)^{{\underline\otimes}\ZZ}$, where
${\underline\otimes}$ denotes the
twisted (w.r.t. root graduation) tensor product~:
$(a{\underline\otimes}b)(c{\underline\otimes}d)
=q^{|b||c|}ac{\underline\otimes}bd$~;
in $U_{q}\np$ the degrees are defined by $|Q_{+}|=-|Q_{-}|=1$. We then have
algebra morphisms $U_{q}\np \ \to \CC[s^{\pm}_{i}]$, defined by $Q_{\pm}\mapsto
s^{\pm}_{i}$, and  $(U_{q}\np)^{{\underline\otimes}\ZZ}\to \CC[s^{\pm}_{i},\,
i\in\ZZ]$ (because $\CC[s_{i}^{\pm}]^{{\underline\otimes\ZZ}}=\CC\langle
s^{\pm}_{i},\, i\in\ZZ\rangle /
(s^{\e}_{i}s^{\e'}_{j}-q^{\e\e'}s^{\e'}_{j}s^{\e}_{i}\, \rm{ if }\, \it i<j)$).

The image of $Q_{\pm}$ by this last morphism is the image of
$\sum\cdots{\underline\otimes}Q_{\pm}{\underline\otimes}1\cdots$, i.e. $\sum
s^{\pm}_{i}$. \ \ $\bull$

The two last relations are obvious. \ \ $\bull$

\nd
{\bf Remark.} The operators $Q_{\pm}$, $K$, defined on the space ${\bf C}
[x_{i}^{\pm1}]$ of arbitrary polynomials by $Q_{\pm}P=
[\sum_{i<0}x_{i}^{\pm1},P]_{q}x_{0}^{\mp1}$, $K= Ad \ x_{0}$
(where $[a,b]_{q}=ab-q^{|a||b|}ba$, and $|\prod_{i\le 0}x_{i}^{\alpha_{i}}|=
\sum_{i\le 0}\alpha_{i}$), satisfy also the relations of Lemma 1.

\section{3.}{Classical results on the lattice.}

{}From Lemma 1 follows that the vector fields $Q_{\pm}^{cl}=
\mp\sum_{j<0}e^{\pm\xi_{j}}(
{\partial\over{\partial\xi_{j}}}+2\sum_{s<j}{\partial\over{\partial\xi_{s}}}
)$, acting on ${\bf C}[e^{\pm\xi_{i}},i<0]$, satisfy the usual affine $sl_{2}$
Serre relations. Let $\sigma$ be the automorphism of
${\bf C}[e^{\pm\xi_{i}},i<0]$ defined by $\sigma(e^{\pm\xi_{i}})=e^{\mp\xi_{i}}
$. Then $\sigma_{*}Q_{\pm}^{cl}=Q_{\mp}^{cl}$ ($\sigma_{*}$ of a vector field
denotes its conjugation by $\sigma$.) So, $\sigma_{*}(Q_{+}^{cl}+Q_{-}^{cl})
=Q_{+}^{cl}+Q_{-}^{cl}$. Similarly, $\sigma_{*}([Q_{+}^{cl},Q_{-}^{cl}])=
-[Q_{+}^{cl},Q_{-}^{cl}]$; posing as in 1, $h(1)=[Q_{+}^{cl},Q_{-}^{cl}]$,
$h(i+1)=[Q_{+}^{cl}, [Q_{-}^{cl}, h(i)]] $, we show by induction that
$\sigma_{*}h(i)=-h(i)$; if it is true for $h(i)$ then $\sigma_{*}h(i+1)=
[Q_{-}^{cl}, [Q_{+}^{cl},- h(i)]]=-h(i+1) $ (by Jacobi identity and
$[h(i),h(1)]=0$). Then $\sigma_{*}[Q_{+}^{cl}-Q_{-}^{cl}, h(i)]=
[Q_{-}^{cl}-Q_{+}^{cl}, -h(i)]$
and so $[Q_{+}^{cl}-Q_{-}^{cl}, h(i)]$ is $\sigma$-invariant. In conclusion,
all
vectors fields of the subalgebra $a\subset\widehat{n}_{+}$, spanned
by
$Q_{+}^{cl}+Q_{-}^{cl}$, and the $[Q_{+}^{cl}-Q_{-}^{cl}, h(i)]$, $i\ge 1$,
are $\sigma$-invariant.

Note that if the vector field $X=\sum_{i<0}X(\xi_{j}){\partial\over{
\partial \xi_{i}}}$ is $\sigma$-invariant, we have $X_{i}(-\xi_{j})=-X_{i}
(\xi_{j})$, so $X_{i}(0)=0$. Let then $\e : {{\bf C}}[e^{\pm\xi_{i}},i<0]
\to {\bf C}$ be the map of evaluation at $\xi_{i}=0$. We have showed that
$\e (xP)=0$, if $x\in a$, $P\in {\bf C}[e^{\pm\xi_{i}},i<0]$, and so the
pairing
$$
U\widehat{n}_{+}\times {{\bf C}}[e^{\pm\xi_{i}},i<0] \to {\bf C},
$$
$$
(T, P)\mapsto\e(TP)
$$
factors through $({{\bf C}}\otimes_{U a}U\widehat{n}_{+})\times
{{\bf C}}[e^{\pm\xi_{i}},i<0]$.

Let us now show that the resulting morphism of $\widehat{n}_{+}$-modules
${{\bf C}}[e^{\pm\xi_{i}},i<0]\to ({{\bf C}}\otimes_{U a}U\widehat{n}_{+})^{*}
$ is an injection. For this, it is enough to show that the Lie algebra
generated by $Q_{+}^{cl}$ and $Q_{-}^{cl}$ contains vector fields
$X^{(n)}=\sum_{k\ge 1}X_{k}^{(n)}(\xi_{-1},\cdots, \xi_{-k})
{\partial\over{\partial \xi_{-k}}}
$ with $X_{k}^{(n)}(0)=0$ for $k<n$,  $X_{n}^{(n)}(0)\not= 0$ for any $n\ge 1$.

We can take $X^{(1)}=Q_{+}^{cl}$, and $X^{(n+1)}=[Q_{+}^{cl}+Q_{-}^{cl},
X^{(n)}]-2X^{(n)}$. By combinations of products of the $X^{(n)}$, it is then
possible to construct in the algebra generated by $Q_{+}^{cl}$  and
$Q_{-}^{cl}$, differential operators of the form $\sum f_{\alpha_{1},\cdots
\alpha_{N}}(\xi)
({\partial\over{\partial\xi_{-1}}})^{\alpha_{1}}\cdots
({\partial\over{\partial\xi_{-N}}})^{\alpha_{N}}
$ + left ideal generated by ${\partial\over{\partial\xi_{-N-k}}}$, $k\ge 1$;
with $f_{\alpha_{1},\cdots,\alpha_{N}}(0)=\delta_{
\alpha_{1},\cdots,\alpha_{N}; \beta_{1},\cdots, \beta_{N}
}$, for any fixed $N\ge 1$ and  $\beta_{1},\cdots, \beta_{N}\ge 0$. Then, any
non zero combination $\sum_{\gamma}\lambda_{\gamma}e^{\sum_{i=1}^{N}\gamma_{i}
\xi_{-i}}$ will have non zero pairing with a combination of the operators
constructed above. We have thus showed:
\proclaim{Proposition 1} The pairing defined above between $U\widehat{n}_{+}$
and ${\bf C}[e^{\pm\xi_{i}},i<0]$ defines an injection of the latter space in
the space of formal series at the origin of $\widehat{N}_{+}/A$, which is an
algebra and $\widehat{n}_{+}$-module morphism.
\endproclaim
Remark that the image of this injection does not contain ${\bf C}[
\widehat{N}_{+}/A]$, because the latter space contains an element ($x_{1}$,
or $\phi$ in the formalism of [FF1]) such that $Q_{+}^{cl}x_{1}=Q_{-}^{cl}x_{1}
=1$, and such an element does not exist in $\CC[e^{\pm\xi_{i}},i<0]$.

\section{4. Quantum principal commutative subalgebra.}{}

Let us assume $q$ to be generic and denote by $U_{q}\widehat{b}_{+}$
the algebra generated by $K$, $Q_{\pm}$, subject to the relations of Lemma 1;
$U_{q}\widehat{b}_{+}$ is a Borel subalgebra of the full quantum algebra
$U_{q}\widehat{sl}_{2}$
(at level zero).
Denoting by $U_{q}\widehat{b}_{-}$ the opposite Borel subalgebra, we then have
an algebra injection $U_{q}\widehat{b}_{+}\hookrightarrow
(U_{q}\widehat{b}_{-})^{*}
$ ([D]). The coordinate ring corresponding to $U_{q}\widehat{b}_{-}$, denoted
${\bf C}[\widehat{B}_{-}]_{q}$,
is the algebra generated by $t_{ij;n}$,
$i,j=1,2$, $n\ge 0$, with $t_{12;0}=0$ and relations
$$
R(\lambda, \mu)T^{(1)}(\lambda)T^{(2)}(\mu)
=T^{(2)}(\mu)T^{(1)}(\lambda)R(\lambda, \mu), \, \, \rm{and}  \, \,
{\mathop{det}}_{q}T(\la)=1
$$
(see [T]), where
$T(\lambda)=(t_{ij}(\lambda))_{1\le i,j\le 2}
=(\sum_{n\ge 0}t_{ij;n}\lambda^{n})_{1\le i,j\le 2}
$, and $R(\lambda,\mu)$ is proportional to the $R$-matrix of [J]:
$$
R(\la,\mu)={{1+q^{1/2}}\over 2}(\la-\mu q^{1/2})
+{{1-q^{1/2}}\over 2}(\la+\mu q^{1/2})h\otimes h
-(q-1)(\la f\otimes e+\mu e\otimes f),
$$
with $h=\pmatrix{ 1 & 0\cr 0 & -1\cr}$, $e=\pmatrix{ 0 & 1\cr
0 & 0\cr}$, $f=\pmatrix{ 0 & 0\cr
1 & 0\cr}$.
We will show:
\proclaim{Lemma 3} The injection $U_{q}\widehat{b}_{+}\hookrightarrow
(U_{q}\widehat{b}_{-})^{*}$ induces an algebra isomorphism between
$U_{q}\widehat{b}_{+}$ and ${\bf C}[\widehat{B}_{-}]_{q}$.\endproclaim

{\it Proof. } The pairing between ${\bf C}[\widehat{B}_{-}]_{q}$ and
$U_{q}\widehat{b}_{-}$ is given by $\langle t_{ij;n}, x\rangle
=\ {\rm res}_{\la=\infty}\la^{n-1}\langle i|\pi\circ T_{\la}(x)|j\rangle
d\la$, $|1\rangle=\pmatrix{ 1\cr 0\cr}$, $|2\rangle=\pmatrix{ 0\cr 1
\cr}$, is the notations of [LSS], app. This enables to identify
$\eta_{1}$, $\eta_{2}$, $e^{\hbar\xi_{1}}$ of \it loc. cit., \rm 7
  with $t_{12;1}$, $t_{21;0}$, $t_{11; 0}=t_{22; 0}^{-1}$, respectively.
The statement can be seen inductively from the relations defining
${\bf C}[\widehat{B}_{-}]_{q}$ (for example, the relation $(1-q)\la(
t_{22}(\la)t_{11}(\mu)-t_{22}(\mu)t_{11}(\la))=
q^{{1\over 2}}(\la-\mu)[t_{21}(\mu),t_{12}(\la)]$ gives
$(1-q)(t_{22;1}t_{11;0}-t_{22;0}t_{11;1})=q^{{1\over 2}}[t_{21;0},t_{12;1}]$,
and the
determinant relation gives $\alpha t_{11;0}t_{22;1}+\beta t_{11;1}t_{22;0}
=t_{21;0}t_{12;1}$, with $\alpha,\beta\to 0$ when $q\to 1$, so combinations of
these relations give $t_{11;1}$ and $t_{22;1}$ in terms of the generators).
$\bull$

Remark the difference with the classical situation, where ${\bf C}[
\widehat{B}_{-}]_{q}$ is not finitely generated; though as Poisson algebra it
is generated by $t_{11;0}$, $t_{12;1}$ and $t_{21;0}$. %notation correcte
Note also that $U_{q}\widehat{n}_{+}$
can be considered as possessing two classical limits, one being the non
commutative algebra $U\np$ and the other being the Poisson algebra generated by
$Q_{+},\, Q_{-}$ and relations $\{Q_{\pm},\{Q_{\pm},\{Q_{\pm},Q_{\mp}\}\}\}=
Q_{\pm}^{2}\{Q_{\pm},Q_{\mp}\}
$ (it is the limit for $\hbar\to 0$ of the $q$-Serre relations, with
$\{a,b\}=\lim_{\hbar\to 0}{1\over\hbar}[a,b]$) and
$q=e^{\hbar}$); these relations are satisfied in particular for
$Q_{\pm}=\int_{S^{1}}e^{\pm\varphi}$, $\varphi$ classical free field.

We will now construct a quantum analogue of the principal commutative
subalgebra of $\widehat{sl}_{2}$.

\proclaim{Proposition 2} For $u(\la)=d,\Lambda d$ ($\Lambda=e+\la f$, $d$ any
diagonal matrix, independant of $\la$), the set of coefficients of $\la^{k}$
($k\ge 0$) in $\tr u(\la)T(\la)$ forms a commutative family in
${\bf C}[\widehat{B}_{+}]_{q}$. For $u=\Lambda\pmatrix{ d_{1} & 0\cr
0&d_{2}\cr}$, with $d_{1}d_{2}\ne 0$, the classical limit of the
corresponding family in $U_{q}\wh{b}_{+}$ is the subalgebra of $U\wh{b}_{+}$
generated by the principal commutative subalgebra spanned by
$d_{2}e(i)+d_{1}f(i+1)$, $i\ge 0$ ($e(0)$ and $e(1)$ denote the elements of
$U\wh{b}_{+}$ corresponding to $\eta_{1}$, $\eta_{2}$ of [LSS], and
$e(i+1)=[e(0),[f(1),e(i)]]$, $f(i+1)=[f(1),[e(0),f(i)]]$ for $i\ge 1$).
\endproclaim

{\it Proof. } For the first part, we first check that $u(\la)\otimes u(\mu)$
commutes with $R(\la,\mu)$ in the two cases. Then
$$\eqalign{
\tr u(\la)T(\la)\tr u(\mu)T(\mu) & =
\tr u(\la)\otimes u(\mu)T^{(1)}(\la)T^{(2)}(\mu)
\cr
& = \tr u(\la)\otimes
u(\mu)R(\la,\mu)^{-1}T^{(2)}(\mu)T^{(1)}(\la)R(\la,\mu) \cr
& = \tr u(\la)\otimes u(\mu)T^{(2)}(\mu)T^{(1)}(\la) =
\tr  u(\mu)T(\mu)     \tr u(\la)T(\la).
}$$
To prove the second part, we first observe that the enveloping algebra of the
principal commutative subalgebra is exactly the centralizer in
$U\widehat{b}_{+}$ of
$d_{2}e(0)+d_{1} f(1)$. This can be seen in the associated graded algebra
$\CC[\widehat{b}^{*}_{+}]$~; in the basis $z_{i}=$ image of $h(i), i\geq
0\ [h(0)$ is the
element of $U\widehat{b}_{+}$ corresponding to $\xi_{1}$ of [LSS], and
$h(i+1)= [e(0),
[f(1), h(i)]]$ for $i\geq 0], x_{i}=$ image of $d_{2}e(i)+d_{1}f(i+1), i \geq
0$, and $y_{i}=$ image of $d_{2}e_{j}-d_{1}f_{i+1}, i\geq 0$, the Poisson
bracket with $x_{0}$ is the vector field $\sum_{i\geq
0}2(-1)^{i+1}y_{i}{\partial\over \partial h_{i}}+ (-1)^{i+1}d_{1}d_{2} h_{i+1}
{\partial\over \partial y_{i}}$~; ordering the basis as
$(z_{i},h_{0},y_{0},h_{1},y_{1},\cdots)$, we see that the only polynomials in
$z_{i}, h_{i},y_{i}$ in the kernel of the vector field are those depending on
$z_{i}$ only.

The image in $U\widehat{b}_{+}$ (by the specialisation $q=1$) of the
commutative
subalgebra of $U_{q}\widehat{b}_{+}$ corresponding to $\tr u(\la) T(\la)$ is
commutative, and it contains $d_{2}e(0)+d_{1} f(1)$. It remains to see that the
subalgebra generated by $\tr u (\la) T(\la)$ is maximal as a commutative
subalgebra of $\CC [\widehat{B}_{+}]_{q}$. We will show it for the
corresponding Poisson subalgebra of $\CC[\widehat{B}_{+}]$. Denote
$T(\la)=(t_{ij} (\la))= \pmatrix{ a(\la) & c(\la)\cr b(\la) & d(\la)\cr}$, with
$a(\la)= \sum_{n\geq 0} a_{n} \la^{n}$, etc. $(b_{0}=0)$. The Poisson brackets
between the variables $a_{n},b_{n},\cdots, $ are given by $\{ T(\la), \otimes
T(\mu) \} = [r(\la ,\mu), T(\la)\otimes T(\mu)]$, with $r(\la,\mu)= {1\over 2}
{\la +\mu\over \la - \mu} h \otimes h + {2\la \over \la -\mu} f \otimes e +
{2\mu \over \la -\mu} e \otimes f$ (trigonometric $r$-matrix). Let us prove
that the polynomials in $a_{n},b_{n},\cdots,$ commuting with $b_{1}-c_{0}$ (to
simplify; the proof with $d_{2}b_{1}+d_{1} c_{0}$ instead is similar
\footnote{*}{(here $d_{i}$ denote the coefficients of the diagonal matrix)}),
are exactly the polynomials in $b_{n+1} -c_{n}(n\geq 0)$. By specializing for
$\mu =0$ the formulas for $\{ a(\la), b(\mu)\},\cdots, $ we get $\{
b_{1}-c_{0}, a(\la)\}= (b_{1}+c_{0})a(\la)-2 a_{0} ( {b(\la)\over \la} +
c(\la))$,
$\{ b_{1}-c_{0}, {b(\la)\over \la}+ c (\la)\} = {4\over \la}
(d_{0}a(\la)-a_{0}d(\la)) , \{ b_{1}-c_{0}, {b(\la)\over \la}- c(\la)\}=0\ .$

So,
$
\{ b_{1}-c_{0}, a_{n}\} = (b_{1}+c_{0})a_{n}-2 a_{0} (b_{n+1}+c_{n}),
\{ b_{1}-c_{0}, b_{n+1}+c_{n}\}= 4 (d_{0}a_{n+1} -a_{0} d_{n+1}),
\{ b_{1}-c_{0}, b_{n+1} -c_{n}\}=0\ ,
$
for $n\geq 0$.

{}From $\det T(\la)=1$, we obtain $d_{0} a_{n+1} - a_{0} d_{n+1} = 2 d_{0}
a_{n+1} + c_{0} b_{n+1} + c_{n} b_{1}+ $ terms in $b_{i} , i\leq n, c_{i} , i
\leq n-1 , a_{i} , i \leq n$. Note that $c_{0} b_{n+1} + c_{n} b_{1} =
{1\over 2} [(c_{0} + b_{1} )(c_{n} + b_{n+1}) + (b_{1} -c_{0})
(b_{n+1}-c_{n})]$. Pose
for $i\geq 0\ ,\ z_{i}= b_{i+1}-c_{i}$ and $x_{i} = b_{i+1} +c_{i}$. The
polynomials in $a_{n}, b_{n},c_{n},d_{n}$ are then the polynomials in
$a^{-1}_{0}, a_{i},x_{i},z_{i} (i \geq 0)$. In this basis the vector field
$\partial =\{ b_{1}-c_{0}, \ \}$ is expressed by $\partial(a_{n})=x_{0} a_{n}-
2a_{0} x_{n}$, $\partial(x_{n}) = 2d_{0} a_{n+1} + {1\over 2} x_{0} x_{n} +$
terms in $a_{i} , i\leq n, x_{i}, i \leq n-1, z_{i}$, and $\partial (z_{n})=0$.
The same argument as above can then be applied, with ordering $(z_{i}, a^{\pm
1}_{0} , x_{1}, a_{1} , x_{2},\cdots)$. Explicitly, let $P(z_{i},a_{i},x_{i})$
be a polynomial, and $x_{i}$ (or $a_{i}$) be the greatest terms on which $P$
depends non trivially~; then the terms in $d_{0}a_{i+1}$ (resp. $a_{0} x_{i})$
of $\partial P$ will be $2 {\partial P\over \partial x_{i}} d_{0}a_{i+1}$
(resp. $-2 {\partial P\over \partial a_{i}}a_{0}x_{i}$ if $i\not= 0$, and
$- {\partial P\over \partial a_{0}} a_{0} x_{0}$ else)~; $\partial P=0$ implies
then ${\partial P\over \partial x_{i}}=0$, (resp. ${\partial P\over \partial
a_{0}} =0)$, contradiction.

As a by-product of this proof, we obtain~:

\proclaim{Corollary} For $q$ generic or $q=1$, the centralizer of $Q_{+}-Q_{-}$
forms a maximal commutative subalgebra of $U_{q}\widehat{b}_{+}$.
\endproclaim
\noindent
{\it Proof. } For $q=1$, it is the first part of the proof above. For $q$
generic, we translate the statement for $\CC[\widehat{B}_{+}]_{q}$, and use the
limit $q\to 1$ and the second part of the proof above. $\bull$

We will call this subalgebra of $U_{q} \widehat{b}_{+}$ its \sl quantum
principal commutative subalgebra \rm and denote it $U_{q}a$; note that
$U_{q}a$ is not a Hopf
subalgebra of $U_{q}\widehat{b}_{+}$ ($a$ is already not a subbialgebra of
$\widehat{b}_{+})$.

\section{5.}{Realisation of $U_{q}a$ in $q$-commuting variables.}

Let us go back to the setting of Lemma 2. It gives an algebra morphism
$U_{q}\widehat{n}_{+} \to \CC [s^{\pm}_{i}]$, and also by composition
$U_{q}\widehat{n}_{+} \to \CC [s^{\pm}_{i}]/(s_i^{\pm} s_i^{\mp}=q_i)$, $q_i$
being invertible scalars.
 Let us describe the image of $U_{q}a$
by this morphism. For this we need to construct the morphism $\CC[\widehat
{B}_{+}]_{q} \to \CC [k, s^{\pm}_{i}]$ deduced from $U_{q}\widehat{b}_{+}
\to \CC
[k, s^{\pm} _{i}]$ by the isomorphism $U_{q} \widehat{b}_{+} \simeq \CC
[\widehat{B}_{+}]$ ($k$ is an additional variable, with $k s^{\pm}_{i} =
q^{\mp{1\over 2}}s^{\pm}_{i}k$, and we prolongate $U_{q} \widehat n_{+}\to \CC
[s_{i}^{\pm}]$ by $K\mapsto k$). From Lemma 3, we see that it is defined by
$t_{11;0} \mapsto k, t_{22;0} \mapsto k^{-1}, t_{12;1}\mapsto \sum s^{+}_{i}\
,\
t_{21;0} \mapsto \sum s^{-}_{i}$.

Let $k_{i}, u^{\pm}_{i}$ be auxiliary
variables, with $k_{i}u^{\pm}_{i} = q^{\mp{1\over 2}} u^{\pm}_{i} k_{i}$, other
relations being commutation relations, and $\prod k_{i}=k$, $\prod_{j<i} k^{\pm
1}_{j} u^{\mp}_{i} \prod_{j>i} k_{j}^{\mp 1} = s^{\mp}_{i}$. Note that we may
impose that $u_i^{\pm}u_i^{\mp}=q_i$. Following Volkov
([Vo]), we remark that the matrices ${1\over{(1-\lambda q q_i)^{1/2}}}
\pmatrix{k_{i} & u^{-}_{i}\cr \la
u^{+}_{i} & k_{i}^{-1}}$, and hence also the matrix $T'(\la)=
\prod^{+\infty}_{i=-\infty} {1\over{(1-\lambda q q_i)^{1/2}}}
\pmatrix{k_{i} & u^{-}_{i}\cr \la
u^{+}_{i} & k_{i}^{-1}}$, satisfy the relations $R(\la
,\mu)T'(\la)^{(1)}T'(\mu)^{(2)} = T'(\mu)^{(2)} T'(\la)^{(1)} R(\la,\mu)$,
$\det_q T'(\lambda)=1$.
Denote $T'(\la)= (t'_{ij}(\la)), t'_{ij}(\la)= \sum_{n\geq 0} t'_{ij;n}
\la^{n}$. The mapping from $\CC[\widehat{B}_{+}]$ to $\CC[k, s^{\pm}_{i}]$,
sending $t_{ij;n}$ to $t'_{ij;n}$ thus extends to an algebra morphism~; since
$t'_{11;0}, t'_{22;0}, t'_{12;1}$ and $t'_{21;0}$ are respectively $k,k^{-1},
\sum s^{+}_{i}, \sum s^{-}_{i}$, this morphism is the desired composition
$\CC[\widehat{B}_{+}]_{q} \simeq U_{q}\widehat b_{+}\to \CC[k,s^{\pm}_{i}]$.

The image of $U_{q}a$ is then generated by
$$
t'_{21;n} - t'_{12;n+1} = \sum_{i_1 < \cdots < i_{2n+1}}\left( \prod_{i< i_{1}}
k_{i}\right) u^{+}_{i_{1}} \left(\prod_{i_{1}<i<i_{2}} k_{i}^{-1} \right)
u^{-1}_{i_{2}}\cdots
u^{+}_{i_{2n+1}} \left(\prod_{i>i_{2n+1}} k_{i}\right)
$$
$$
- \left(\prod_{i < i_{1}}k_{i}^{-1}\right) u_{i_{1}}^{-} \left(
\prod_{i_{1}<i<i_{2}}
k_{i}\right) u^{+}_{i_{2}}\cdots u^{-}_{i_{2n+1}}\left(
\prod_{i> i_{2n+1}} k_{i}^{-1}\right)
$$
$$+\sum_{p<n}\rm scalars\,  \it ( \rm analogous\, \, expression\, \, with \, \,
\it n \, \,
\rm replaced \, \, by \, \, \it p)
$$

for $n\geq 0$, which can be written
$$
t'_{21;n} - t'_{12;n+1} = \sum_{i_1< \cdots < i_{2n+1}}
s^{+}_{i_{1}} s^{-}_{i_{2}}\cdots s^{+}_{i_{2n+1}} - s^{-}_{i_{1}}s^{+}_{i_{2}}
\cdots s^{-}_{i_{2n+1}}
$$
$$+\sum_{p<n}\rm scalars\,  \it ( \rm analogous\, \, expression\, \, with \, \,
\it n \, \,
\rm replaced \, \, by \, \, \it p).
$$
We have proved~:

\proclaim{Lemma 4} The image of the principal commutative subalgebra of
$U_{q}\widehat n_{+}$, by the mapping defined in Lemma 2, is the subalgebra of
$\CC[s^{\pm}_{i}]$ generated by
$$
\sum_{i_1 < \cdots < i_{2n+1}}s^{+}_{i_{1}}s^{-}_{i_{2}}\cdots
s^{+}_{i_{2n+1}} - s^{-}_{i_{1}}s^{+}_{i_{2}}\cdots s^{-}_{i_{2n+1}}\ ,\
\qqbox{for} n\geq 0\ .
$$
\endproclaim

Note that in the case where there is only a finite number $N$ of $s^{\pm}_{i}$
the image of $U_{q}a$ is finitely generated (the sums vanish for $n\geq
[{N+1\over 2}]$). One may think that the elements $t'_{21;n}- t'_{12;n+1}$,
for $n\geq [{N+1\over 2}]$, generate the kernel of the morphism $U_{q}\widehat
n_{+}\to \CC [s^{\pm}_{i}]$, and that this morphism is injective if there is an
infinite number of $s^{\pm}_{i}$.

\section{6.}{The pairing between $U_{q}\widehat n_{+}$ and the lattice KdV
variables.}

Recall that in sect. 2, $K = Ad\,  x_{0} =
q^{-\sum_{s<0}{\partial\over \partial
\xi_{s}}}=\t '_{0}$ (posing $\xi_{0} = 0$). The arguments of sect. 2 show that
the operators $\overline Q_{+} =-\sum_{j<0} \t_{j} + \sum_{j\leq 0} \t '_{j} =
Q_{+}
+ \t '_{0}$, and $\overline Q_{-} = - \sum_{j<0} \t^{-}_{j} + \sum_{j\leq 0}
\t'^{-}_{j'}$ (where $\t '^{-}_{0} = \t '^{-1}_{0}$) satisfy the
$q$-Serre relations.

Let us consider the algebra mapping $\e : \CC [ e^{\pm \xi_{i}}]\to \CC$,
defined by $e^{\pm \xi_{i}} \mapsto 1$. We can compose it with the action of
$U_{q} \widehat n_{+}$ (by $\overline Q_{+}$ and $\overline Q_{-}$) on
$\CC[e^{\pm\xi_{i}}]$, and obtain a pairing between $U_{q}\widehat n_{+}$ and
$\CC[e^{\pm\xi_{i}}]$.

Let us show that for any polynomial $P\in \CC[ e^{\pm\xi_{i}}]$, and $n\geq 0$,
$\e((t_{21;n} -t_{12;n+1})P)=0$. Ordering the $\t_{i} ,\t'_{j}$ by $(\t_{-1},
\t_{-2},\cdots, \cdots, \t '_{-1},\t'_{0})$, Lemma 4 shows that
$$
(t_{21;n}-t_{12;n+1} )P = \left(\sum_{i_{1}<\cdots <
i_{2n+1}}\f^{+}_{i_{1}}\f^{-}_{i_{2}} \cdots \f^{+}_{i_{2n+1}}
-\sum_{i_1<\cdots <
i_{2n+1}} \f^{-}_{i_{1}} \f^{+}_{i_{2}}\cdots \f^{-}_{i_{2n+1}}\right) P\ ,
$$
$\f^{\pm}_{i}$ is the list $(\t^{\pm}_{-1},\cdots,\t'^{\pm}_{0})$. We split
each of these sums in two parts~: the terms such that for some $\a$,
$\f_{i_{1}}= \t_{\a}$ \footnote{*}{we note also $\f^{+}_{i} =\f_{i}$,
$\t^{+}_{i} =\t_{i}$},
and $\f_{i_{2n+1}} = \t'_{\a+1}$ and the other terms for the first sum, and the
terms such that $\f^{-}_{i_{1}} =\t^{-}_{\a}$ and $\f_{i_{2n+1}}^-= \t
'^{-}_{\a+1}$ and the other terms for the second. We can define a bijection
between the sets of remaining terms in the following way~: to
$\f_{i_{1}}\f'^{-}_{i_{2}}\cdots \f_{i_{2n+1}}$, with $\f_{i_{1}}=\t_{\a}$
and $\f_{i_{2n+1}}=\t'_{\b+1}$ we associate $\f^{-}_{i_{2}} \f_{i_{3}} \cdots
\f_{i_{2n+1}} \t'^{-}_{\a+1}$ if $\a > \b$, and
$\t^{-}_{\b}\f_{i_{1}}\f^{-}_{i_{2}}\cdots \f^{-}_{i_{2n}}$ if $\a < \b$. In
both cases, $\e((\f_{i_{1}} \f_{i_{2}}^{\prime -}\cdots \f_{i_{2n+1}}-$ its
associated
term) $P)=0$. Indeed, in the first case $\f_{i_{1}}= e^{\xi_{\a}}
q^{\sum_{s\leq\a} {\partial\over \partial \xi_{s}}}$, and $\t'^{-}_{\a+1}
=e^{-\xi_{\a+1}}q^{\sum_{s\leq \a} {\partial\over \partial\xi_{s}}}$. $\a+1$ is
larger than all indexes occuring in $\f^{-}_{i_{2}}\cdots \f_{i_{2n+1}}$ so
$e^{-\xi_{\a+1}}$ can be translated to the left (in the expression
$\f^{-}_{i_{2}}\cdots \f_{i_{2n+1}} \t'^{-}_{\a+1}$) without changing the
result, and there is also no correction due to the transport of
$q^{\Si_{s\leq\a}{\partial\over \partial\xi_{s}}}$ to the left, because it has
to cross the same number of $e^{\xi_{i}}$ and $e^{-\xi_{j}}$, with all theses
$i$ and $j$ less than $\a$. In conclusion, we can identify $\f_{i_{1}}
\f'^{-}_{i_{2}}\cdots \f_{i_{2n+1}}$ with $e^{\xi_{\a}+\xi_{\a+1}}$.(its
associated term). Similarly, in case $\a < \b$, $\f_{i_{1}}
\f'^{-}_{i_{2}}\cdots
\f_{i_{2n+1}}$ is identified with $e^{-\xi_{\b}-\xi_{\b+1}}$.(its associated
term) so if $\a\not= \b\ ,\ \e((\f_{i_{1}} \f'^{-}_{i_{2}}\cdots \f_{i_{2n+1}}
-$ associated term)$P)=0$.

For the first parts of the sums, we divide them in partial sums $\Si_{\a}$,
with $\f_{i_{1}} =\t_{\a}$ and $\f_{i_{2n+1}}=\t'_{\a+1}$ (resp.
$\f'^{-}_{i}=\t^{-}_{\a}$ and $\f^{-}_{i_{2n+1}}=\t'^{-}_{\a+1}$). Then
$\t_{\a} \f^{-}_{i_{2}} \f_{i_{3}}\cdots \f^{-}_{i_{2n}} \t'_{\a+1}=
e^{\xi_{\a+1} + \xi_{\a+1}} \f^{-}_{i_{2}}\f_{i_{3}} \cdots \f^{-}_{i_{2n}}$,
and $\t^{-}_{\a}\f_{i_{2}} \f^{-}_{i_{3}}\cdots \f_{i_{2n}} \t'^{-}_{\a+1} =
e^{-\xi_{\a}-\xi_{\a+1}}\f_{i_{2}}\f^{-}_{i_{3}}\cdots
\f_{i_{2n-1}}^{-}\f_{i_{2n}}$. So $\e (\Si_{\a}.P)=$
$$
=\e\left[ (\sum_{i_1(\a)< i_{2}\cdots < i_{2n}< i_{2n+1}(\a)}\f^{-}_{i_{2}}
\f_{i_{3}}\cdots \f^{-}_{i_{2n}}- \f _{i_{2}}\f^{-}_{i_{3}}\cdots
\f_{i_{2n}}) P\right]\  ;
$$
this is an expression of the same type that the expression we started with,
with smaller degree. So we can use an induction argument to show that these
expressions vanish.

So $\e((t_{21;n} - t_{12;n+1})P)=0$ as claimed. And we can state the first
part of~:

\proclaim{Theorem} The pairing between $U_{q}\widehat n_{+}$ and
$\CC[e^{\pm\xi_{i}}]$, given by
$$
U_{q}\widehat n_{+}\times \CC [e^{\pm\xi_{i}}]\to
\CC[e^{\pm\xi_{i}}]\buildrel\e\over \longrightarrow \CC\ ,
$$
where the first map is the action of $U_{q}\widehat n_{+}$ on
$\CC[e^{\pm\xi_{i}}]$, factors through a pairing
$$
(\CC \otimes_{U_{q}a}U_{q}\widehat n_{+})\times \CC [e^{\pm\xi_{i}}]\to \CC\ ,
$$
which induces an injection of $U_{q}\widehat n_{+}$-modules
$\CC[e^{\pm\xi_{i}}]\hookrightarrow (\CC \otimes_{U_{q}a} U_{q} \widehat
n_{+})^{*}$.
\endproclaim

To prove the injection statement, we note that the classical limit of the
operator
$\overline Q_{\pm}$ is $\overline Q^{c\ell}_{\pm}=
Q_{\pm}^{c\ell}+1$. Let $\f$ be a
function on $\widehat N_{+}$ such that $Q_{+}\f = Q_{-}\f =1$~; $\f$ is (up to
an additive constant) the function assigning to $\exp(\a_{0}e(0))\exp
(\b_{1}f(1)) \exp (\a_{1}e(1))$ $\exp (\b_{2}f(2))\cdots \in
\widehat{N}_{+},e^{\a_{0}+\b_{1}}$ (in the notations of prop. 2). Denoting by
$\iota$ the injection $\CC[e^{\pm \xi_{i}}]\to (\CC\otimes_{U a} U \widehat
n_{+})^{*}$ provided by the operators $Q^{c\ell}_{\pm}$, the analogous mapping
$\bar \iota$, provided by $\overline Q_{\pm}^{c\ell}$, will be $\bar\iota
=\f \iota$ (composition
of $\iota$ with the multiplication by $\f$), and so will also be an injection.
Since by [LSS], the family $U_{q}\widehat n_{+}$ is flat at $q=1$ (PBW result),
and by prop. 2, the limit of $U_{q}a$ is $Ua$, the quantum mapping
$\CC[e^{\pm \xi_{i}}]\to (\CC \otimes_{U_{q}a} U_{q}\widehat n_{+})^{*}$ has
for limit the classical mapping $\CC[
e^{\pm\xi_{i}}]\buildrel{\bar\iota}\over\longrightarrow
(\CC \otimes_{U_{q}a}U\widehat n_{+})^{*}$, which is injective, and so is
injective.

\noindent
{\bf Remerciements. } Je voudrais remercier Mmes Harmide et Truc pour la frappe
de ce texte.

\medskip
\noindent
{\bf References}

\bibitem{[D]} V.G. Drinfeld, Quantum Groups, Proc. ICM Berkeley, vol. 1,
798-820 (1988).
\medskip
\bibitem{[F]} B.L. Feigin, Moscow lectures (1992).
\medskip
\bibitem{[FF1]} B.L. Feigin, E. Frenkel, Integrals of motion and quantum
groups, to appear in Proc. CIME summer school ``Integrable systems and quantum
groups'', Springer~; hep-th 9310022.
\medskip
\bibitem{[FF2]} B.L. Feigin, E. Frenkel, Generalized KdV flows and nilpotent
subgroups of affine Kac-Moody groups, hep-th 9311171
\medskip
\bibitem{[J]} M. Jimbo, Quantum $R$-matrix for the generalized Toda system,
Commun. Math. Phys. 102, 537-547 (1986).
\medskip
\bibitem{[KP]} S.V. Kryukov, Ya.P. Pugay, Lattice $W$-algebras and quantum
groups, Landau-93-TMP-5, hep-th 9310154.
\medskip
\bibitem{[LSS]} S. Levendorskii, Y. Soibelman, V. Stukopin, Quantum Weyl group
and universal quantum $R$-matrix for affine Lie algebra $A^{(1)}_{1}$, Lett.
Math. Phys. (1993).
\medskip
\bibitem{[T]} V. Tarasov, Cyclic monodromy matrices of $SL(n)$ trigonometric
$R$-matrices, RIMS-903, hep-th 9211105.
\medskip
\bibitem{[V]} A. Yu. Volkov, Quantum Volterra model, Phys. Lett. A167 (1992),
345-355.
\medskip
\adresse{Centre de Math\'{e}matiques\cr
URA 169 du CNRS\cr
Ecole Polytechnique\cr
91128 Palaiseau FRANCE\cr}

\end